\documentstyle[aps,prl,multicol,epsf]{revtex}
\begin{document}
\widetext

\title{Nature of correlations in the atomic limit \\
       of the boson fermion model}

\author{T.\ Doma\'nski}
\address{Institute of Physics, 
         M.\ Curie Sk\l odowska University,
         20-031 Lublin, Poland}

\date{\today}
%\begin{document}
\maketitle
\draft  

\begin{abstract}
Using the equation of motion technique for Green's functions we 
derive the exact solution of the boson fermion model in the atomic 
limit. Both (fermion and boson) subsystems are characterised 
by the effective three level excitation spectra. We compute
the spectral weights of these states and analyse them in 
detail with respect to all possible parameters.
 
Although in the atomic limit there is no true phase transition, we 
notice that upon decreasing temperature some pairing correlations 
start to appear. Their intensity is found to be proportional to
the depleted amount of the fermion nonbonding state. We notice
that pairing correlations behave in a fashion observed for the
optimally doped and underdoped high $T_{c}$ superconductors.
We try to identify which parameter of the boson fermion model
can possibly correspond to the actual doping level. 
This study clarifies the origin of pairing correlations 
within the boson fermion model and may elucidate how to apply 
it for interpretation of experimental data.
\end{abstract}

\begin{multicols}{2}
 
\narrowtext      
Boson fermion (BF) model describes a system composed of the narrow 
band electrons or holes (fermions) which coexist and interact with 
the local pairs (hard-core bosons) of, for example, bipolaronic
origin \cite{model}. The BF model has been recently intensively 
studied by various methods, such as: the standard mean field 
theory \cite{model}, the perturbative procedure with respect to 
the boson fermion coupling \cite{perturbative}, perturbative
expansion with respect to the kinetic hopping \cite{Domanski98}, 
the dynamical mean field procedure \cite{romano}, the continuous 
canonical transformation \cite{Domanski01}, etc. Apart of studying
the mechanism responsible for superconductivity, there have been 
also investigated the many-body effects which, above $T_c$, lead 
to an appearance of fermion pairs without their long range
coherence. Indeed, three independent procedures 
\cite{perturbative,romano,Domanski01} gave unambiguous 
arguments for the precursor effects, out of which a pseudogap 
is the most transparent one. 

The pseudogap feature gradually builds up upon lowering temperature.  
It is observed in a temperature regime $T^{*} > T > T_{c}$, with
both characteristic temperatures $T^{*}$ and $T_{c}$ depending on
the BF model parameters. Absence of the long range coherence between
pairs is caused by quantum fluctuations of the order parameter 
$\left<c_{i\downarrow}c_{i\uparrow}\right> \equiv \chi_{i}
e^{i\phi_{i}}$. In general, it is hard to distinguish between 
the amplitude $\chi_{i}$ and phase $\phi_{i}$ fluctuations because 
they are convoluted. Intuitively one may expect that phase fluctuations 
would dominate for a dilute concentration of paired fermions, while 
in the opposite limit the amplitude fluctuations take over.
Some analysis along this line was recently discussed in Ref.\ 
\cite{Tripodi_02}. Fluctuation effects were also studied 
for the 2 dimensional (isotropic and anisotropic) BF model 
by Micnas {\em et al} \cite{Micnas_02} using the Kosterlitz 
Thouless theory. Authors reported a noticeable splitting
between $T^{*}$ and $T_{c}$ which considerably increased 
for increasing population of the paired fermions. This 
result supports the above mentioned reasoning. 

In this brief report we show that already on a level of the 
zero-dimensional (atomic limit) physics there is some evidence 
for pairing correlations which gradually increase in strength 
upon lowering temperature. We study such effect on a basis of 
the rigorous solution of the BF model in the atomic limit. 

In our previous paper \cite{Domanski98} we have investigated 
some aspects of the atomic limit solution. The effective fermion 
spectrum was determined there by a direct diagonalisation of 
the Hilbert space. In a current work we rederive the exact 
solution using the equation of motion technique \cite{Zubariev60} 
for Green's functions. Advantage of this method is that it gives 
the spectral weights for the eigenstates expressed in terms of 
the corresponding correlation functions. Of course, diagonalisation
and Green's function method are equivalent and complementary 
to each other.

Hamiltonian of the boson fermion  model can be written as 
$H = \sum_{i,j,\sigma} t_{i,j} c_{i,\sigma}^{\dagger} c_{j,\sigma} 
+ \sum_{i} H_{i}$ where $t_{ij}$ stands for the hopping integral
and the local part $H_{i}$ is given by \cite{model}
\begin{eqnarray}
H_{i} & =  & \varepsilon_{0} \sum_{\sigma} c_{i,\sigma}^{\dagger}
c_{i,\sigma} + E_{0} b_{i}^{\dagger} b_{i} 
\nonumber \\ & + & g \left( b_{i}
c_{i,\uparrow}^{\dagger} c_{i,\downarrow}^{\dagger} + 
b_{i}^{\dagger} c_{i,\downarrow} c_{i,\uparrow} \right)
\;.
\label{local}
\end{eqnarray}
We use here standard notations for the second quantisation
operators of fermion $c_{i,\sigma}$, $c_{i,\sigma}^{\dagger}$ 
and hard core boson $b_{i}$, $b_{i}^{\dagger}$ fields. Site 
energies are correspondingly expressed as $\varepsilon_{0}
=\varepsilon_{f}-\mu$ and $E_{0}=\Delta_{B} - 2 \mu$ where 
a common chemical potential $\mu$ ensures conservation 
of the total charge concentration $n_{tot}= \left< 2b_{i}^{\dagger}
b_{i} + \sum_{\sigma} c_{i,\sigma}^{\dagger}c_{i,\sigma}\right>$. 
Fermion and boson fields are coupled through the exchange 
interaction $gb_{i}c_{i,\uparrow}^{\dagger}c_{i,\downarrow}^{\dagger}
+h.c.$ which can transform a fermion pair into a hard core boson 
and {\em vice versa}. 

In the strict atomic limit $t_{ij}=0$ one needs a solution of 
only the local part (\ref{local}). Let us notice that the hard 
core boson operators obey, in general, the spin $\frac{1}{2}$ 
algebra, characterised by the following commutation rules 
$[b_{i},b_{i}^{\dagger}]=\delta_{ij}(1-2b_{i}^{\dagger}b_{i})$ 
and $[b_{i},b_{j}]=0=[b_{i}^{\dagger},b_{j}^{\dagger}]$. 
For the same site $i=j$ (which is relevant in the atomic limit) 
they simply reduce to the anticommutation relations \cite{Micnas92}. 
We can thus construct the fermionic Green's function 
$\langle \langle A_{i};A_{i}^{\dagger} \rangle\rangle_{\omega}$ 
both for fermions $A_{i}=c_{i\sigma}$ and for hard-core 
bosons $A_{i}=b_{i}$, where we introduced the Fourier
transform of  the retarded Green's function $-i \Theta(t)
\left< \left[ A_{i}(t) , A_{i}^{\dagger}(0) \right]\right>
\equiv \int d\omega e^{i\omega t}\langle \langle A_{i};
A_{i}^{\dagger} \rangle\rangle_{\omega}$. 
% The abbreviation $\langle \langle A_{i};A_{i}^{\dagger} 
% \rangle\rangle_{\omega}$ denotes a Fourier transform of 
% the retarded Green's function $-i \Theta(t)\left< A_{i}(t)
% A_{i}^{\dagger}(0) + A_{i}^{\dagger}(0) A_{i}(t)\right>$.

According to the equation of motion \cite{Zubariev60} 
$\omega \langle\langle A;B \rangle\rangle_{\omega} = 
\langle\{A,B\}\rangle + \langle\langle \left[ A,H \right]
;B\rangle\rangle_{\omega}$ we find the following set of 
coupled equations 
\end{multicols} 

\widetext
\begin{eqnarray}
\left( \omega - \varepsilon_{0} \right) \langle\langle 
c_{i,\uparrow};c_{i,\uparrow}^{\dagger} \rangle\rangle_{\omega} 
& = & 1 + g \langle\langle b_{i}c_{i,\downarrow}^{\dagger};
c_{i,\uparrow}^{\dagger} \rangle\rangle_{\omega} \;,
\\
\left( \omega + \varepsilon_{0} - E_{0} \right) \langle\langle
b_{i} c_{i,\downarrow}^{\dagger};c_{i,\uparrow}^{\dagger}
\rangle\rangle_{\omega} & = &
g \langle\langle (n_{i,\downarrow}^{F} - n_{i}^{B} )^{2} 
c_{i,\uparrow};c_{i,\uparrow}^{\dagger} \rangle\rangle_{\omega} \;,
\\
\left( \omega - E_{0} \right) \langle\langle ( n_{i,\downarrow}
^{F} - n_{i}^{B} )^{2} c_{i,\uparrow}; c_{i,\uparrow}^{\dagger}
\rangle\rangle_{\omega} & = &
\langle (n_{i,\downarrow}^{F} - n_{i}^{B})^{2} \rangle +
g \langle\langle b_{i}c_{i,\downarrow}^{\dagger} ; 
c_{i,\uparrow}^{\dagger} \rangle\rangle_{\omega}  \;,
\end{eqnarray}
where $n_{i,\sigma}^{F} = c_{i,\sigma}^{\dagger}c_{i,\sigma}$ 
and $n_{i}^{B}=b_{i}^{\dagger}b_{i}$. After some algebraic
calculations we determine that these three functions read
\begin{eqnarray}
\langle\langle c_{i,\uparrow}; c_{i,\uparrow}^{\dagger}
\rangle\rangle_{\omega} & = & 
\frac{1 - \langle (n_{i,\downarrow}^{F}-n_{i}^{B})^{2}\rangle }
{\omega-\varepsilon_{0}} + \frac{ \langle (n_{i,\downarrow}^{F} -
n_{i}^{B} )^{2} \rangle \; \left( \omega + \varepsilon_{0}
-E_{0}\right)}{\left( \omega - \varepsilon_{0} \right) \left( \omega
+\varepsilon_{0} - E_{0} \right) - g^{2}} \;,
\label{G_analytic}
\\
\langle\langle b_{i} c_{i,\downarrow}^{\dagger}; 
c_{i,\uparrow}^{\dagger} \rangle\rangle_{\omega}
& = & \langle (n_{i,\downarrow}^{F} -
n_{i}^{B} )^{2} \rangle \; \frac{g} 
{\left( \omega - \varepsilon_{0} \right) \left( \omega
+\varepsilon_{0} - E_{0} \right) - g^{2}} \;, 
\label{bcc_analytic}
\\
\langle\langle ( n_{i,\downarrow}^{F} - n_{i}^{B} )^{2} 
c_{i,\uparrow}; c_{i,\uparrow}^{\dagger}\rangle\rangle_{\omega}
& = & \langle (n_{i,\downarrow}^{F} -
n_{i}^{B} )^{2} \rangle \; \frac{\omega+\varepsilon_{0}-E_{0}} 
{\left( \omega - \varepsilon_{0} \right) \left( \omega
+\varepsilon_{0} - E_{0} \right) - g^{2}} \;. 
\label{3GF_analytic}
\end{eqnarray}
It is convenient to rewrite the single particle Green's 
function in the following way
\begin{eqnarray}
\langle\langle c_{i,\uparrow}; c_{i,\uparrow}^{\dagger}
\rangle\rangle_{\omega} & = & \frac{Z^{F}}{\omega - 
\varepsilon_{0}} + \left( 1 - Z^{F} \right)  \left[
\frac{v^{2}}{\omega-\varepsilon_{+}} + \frac{u^{2}}
{\omega-\varepsilon_{-}} \right] \;,
\label{GF} 
\\
Z^{F} & = & 1 - \langle (n_{i,\downarrow}^{F} 
- n_{i}^{B})^{2} \rangle  \;,
\label{ZF}
\\ 
\varepsilon_{\pm} & = & \frac{E_{0}}{2} \pm \sqrt{ \left(
\varepsilon_{0} - \frac{E_{0}}{2} \right)^{2} + g^{2}} \;,
\label{eps_pm}
\\
v^{2} = 1 - u^{2} & = & \frac{1}{2} \left[ 1 + \frac{
\varepsilon_{0} - \frac{E_{0}}{2}}{\sqrt{ \left( \varepsilon_{0}
- \frac{E_{0}}{2} \right)^{2} + g^{2}}} \right] \;.
\label{vu}
\end{eqnarray}
Another set of coupled equations to determine the hard core 
boson propagator $\langle\langle b_{i} ; b_{i}^{\dagger}
\rangle\rangle_{\omega}$ involves the following Green's
functions
\begin{eqnarray}
\left( \omega - E_{0} \right) \langle\langle b_{i} ; b_{i}^{\dagger}
\rangle\rangle_{\omega} & = & 1 + g \langle\langle c_{i,\downarrow}
c_{i,\uparrow} ; b_{i}^{\dagger} \rangle\rangle_{\omega}  \;,
\\
\left( \omega - 2\varepsilon_{0} \right) \langle\langle
c_{i,\downarrow}c_{i,\uparrow} ; b_{i}^{\dagger} \rangle\rangle
_{\omega} & = & 2 \langle c_{i,\downarrow} c_{i,\uparrow} b_{i}
^{\dagger} \rangle + g \langle\langle b_{i} ; b_{i}^{\dagger}
\rangle\rangle_{\omega} - g \sum_{\sigma} \langle\langle
c_{i,\sigma}^{\dagger}c_{i,\sigma}b_{i} ; b_{i}^{\dagger}
\rangle\rangle_{\omega}  \;,
\\
\left( \omega - E_{0} \right) \sum_{\sigma} \langle\langle
c_{i,\sigma}^{\dagger} c_{i,\sigma} b_{i} ; b_{i}^{\dagger}
\rangle\rangle_{\omega} & = & \langle n_{i,\uparrow}^{F}
+ n_{i,\downarrow}^{F} \rangle \;.
\end{eqnarray}

In analogy to (\ref{GF}) we present the explicit form 
of the single particle Green's function as
\begin{eqnarray}
\langle\langle b_{i};b_{i}^{\dagger} \rangle\rangle_{\omega} 
& = & \frac{Z^{B}}{\omega - E_{0}} + \left( 1 - Z^{B} \right) 
\left[ \frac{u^{2}}{\omega - E_{+}} + \frac{v^{2}}{\omega -
E_{-}} \right]  \;,
\label{GB} \\
Z^{B} & = & \langle (n_{i,\uparrow}^{F} - n_{i,\downarrow}^{F}
)^{2} \rangle \;,
\label{ZB} \\
E_{\pm} & = & \varepsilon_{\pm} + \varepsilon_{0}
\label{E_pm} \;.
\end{eqnarray}

%----------------------------------
\begin{multicols}{2}
\narrowtext
The single particle propagators (\ref{GF}) and (\ref{GB}) are 
both characterised by a three pole structure. One of the poles 
is a remnant of the free nonbonding state ($\varepsilon_{0}$  
for fermions and $E_{0}$ for hard core bosons). The other two 
poles ($\varepsilon_{\pm}$ and $E_{\pm}$) correspond to the bonding 
and antibonding states which arise due to the boson fermion 
interaction. Hamiltonian (\ref{local}) is no longer 
diagonal in the occupation representation $| n_{\uparrow}^{F},
n^{F}_{\downarrow};n^{B}\rangle$ because two eigenvectors contain 
admixture of $|\uparrow,\downarrow;0\rangle$ and $|0,0;1\rangle$ 
\cite{Domanski98}. Loosely speaking, an ability of the system 
to fluctuate between these two states is a measure of pairing 
correlations (we mean the correlations in time, because in the
atomic limit there exist no spatial correlations).

Let us inspect in some detail the spectral weight $Z_{F}$ 
of the nonbonding fermions' state. From (\ref{ZF}) we see 
that $Z^{F}$ is depleted from unity by $\langle 
(n_{i,\downarrow}^{F}-n_{i}^{B})^{2}\rangle = \langle 
n_{i,\downarrow}^{F}\rangle + \langle n_{i}^{B} \rangle 
- \langle 2 n_{i,\downarrow}^{F} n_{i}^{B}\rangle$. 
It means that propagation (in time) of the free fermion 
(with spin $\sigma=\uparrow$) occurs unless: (a) there 
exists another fermion on the same site with the opposite 
spin and simultaneously no hard-core boson is present there, 
(b) there is boson while $\downarrow$ fermion is absent.  
Disappearance of the nonbonding state depends thus on 
fermion and boson concentrations. Role of other factors, 
such as for example temperature, is less evident at this point.

Spectral weight of hard core boson nonbonding state is given by
\begin{equation}
Z^{B} = \langle ( n_{i,\uparrow}^{F} - 
        n_{i,\downarrow}^{F} )^{2} \rangle 
      = \langle n_{i}^{F} \rangle - 2 \langle 
        n_{i}^{pair} \rangle \;,
\label{pairs}
\end{equation}
where $n_{i}^{F}=n_{i,\uparrow}^{F}+n_{i,\downarrow}^{F}$ counts
the total number of fermions on site $i$, while $n_{i}^{pair}$ 
counts only the doubly occupied fermion states 
$n_{i}^{pair} \equiv c_{i,\uparrow}^{\dagger}
c_{i,\downarrow}^{\dagger}c_{i,\downarrow}c_{i,\uparrow}$. 
The hard core boson can safely exist in a free (nonbonding) 
state when there are only single fermions present on the same 
site. The more fermions are paired, the less spectral weight 
is left for a free hard core boson.

We can express the spectral weights $Z^{F}$ and $Z^{B}$ explicitly 
via the concentrations $n^{F}\equiv\sum_{\sigma}\langle 
c_{i,\sigma}^{\dagger} c_{i,\sigma}\rangle$, $n^{B}\equiv
\langle b_{i}^{\dagger}b_{i}\rangle$ and through such parameters
as temperature $T$ and $\Delta_{B}$. From a general relation 
\cite{Zubariev60} $\langle AB \rangle = -\frac{1}{\pi}
\int d\omega f(\omega) \mbox{Imag} \langle \langle B;A 
\rangle\rangle_{\omega+i\eta}$ we obtain 
\begin{eqnarray}
Z^{F} & = & \frac{n^{F}-\left[ v^{2} f(\varepsilon_{+})+
u^{2} f(\varepsilon_{-})\right]}{f(\varepsilon_{0})-\left[ 
v^{2} f(\varepsilon_{+})+u^{2} f(\varepsilon_{-})\right]}
\;, \label{ZF_GreenF} \\
Z^{B} & = &  \frac{n^{B}-\left[ u^{2} f(E_{+})+v^{2} 
f(E_{-})\right]}{f(E_{0})-\left[ u^{2} f(E_{+})+v^{2} 
f(E_{-})\right]}
\;, \label{ZB_GreenF}
\end{eqnarray}
where $f(x)=\left[ e^{x\beta}+1\right]^{-1}$ is
the Fermi Dirac distribution and $\beta=1/k_{B}T$. These 
quantities can be computed also from the diagonalized Hamiltonian 
using the Lehmann representation. They are found to be  
\cite{Domanski98}
$Z^{F}=\left[1 + e^{-\beta\varepsilon_{0}} +
e^{-\beta \left( \varepsilon_{0} + E_{0} \right)} +
e^{-\beta \left( 2\varepsilon_{0} + E_{0} \right)}\right]
/\Theta$ ($\Theta=1+2e^{-\beta\varepsilon_{0}}+
2e^{-\beta\left( \varepsilon_{0}+E_{0}\right) } + 
e^{-\beta\left(2\varepsilon_{0}+E_{0}\right)}+
e^{-\beta E_{+}} +e^{-\beta E_{-}}$ is the
partition function) and
$Z^{B}=\left[ 2e^{-\beta\varepsilon_{0}} + 2e^{-\beta
\left( \varepsilon_{0}+E_{0} \right)}\right]/\Theta$. 
These expressions are of course identical with 
(\ref{ZF_GreenF},\ref{ZB_GreenF}).

We explored numerically variation of the spectral weights 
$Z^{F}$, $Z^{B}$ versus temperature $T$ and $\Delta_{B}$ 
for several fixed charge concentrations $n_{tot}=n^F+2n^B$. 
From our analysis it turns out that the most sensitive 
$T$-dependence of these quantities occurs for 
$\varepsilon_{0}+E_{0}=0$ when $n_{tot}=2$. One
can show that 
\begin{equation}
Z^{F}_{|n_{tot}=2} =
\frac{2}{3+\frac{\cosh{\beta\sqrt{(\Delta_{B}/2)^{2}
+g^{2}}}}{\cosh{\left( \beta\Delta_{B}/6 \right) }}} 
= Z^{B}_{|n_{tot}=2} \;.
\end{equation}
which at high temperature approach the asymptotic value 
$lim_{T\rightarrow\infty}Z^{F,B}_{|n_{tot}=2}=0.5$, while 
for $T \longrightarrow 0$ diminish to zero. Figure 1 
illustrates this behaviour.

\begin{figure}
\centerline{\epsfxsize=8cm \epsfbox{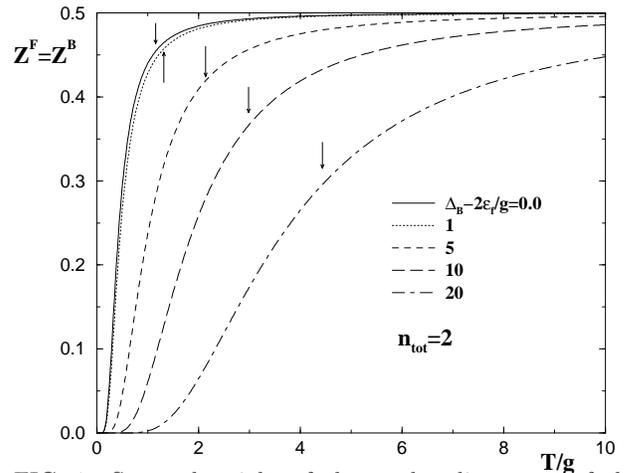}}
\caption{Spectral weight of the nonbonding state of the fermion
and hard core boson subsystems for total charge concentration
$n_{tot}=2$. Main suppresion of the spectral weight of the
nonbonding state occurs near $T^{*}$ (pointed by the arrows)
and depends on the parameter $\Delta_{B}$.}
\end{figure}

In any other case the spectral weights $Z^{F}$, $Z^{B}$
may not vanish in the ground state. They vary within a 
narrower regime signalling that interaction effect is then
less efficient as compared to the case $n_{tot}=2$. Figure 2
shows the spectral weights $Z^{F,B}$ as functions of $n_{tot}$ 
for $\Delta_{B}/2=\varepsilon_{f}$. For fermions we notice that
away of $n_{tot}=2$ the spectral weight $Z^{F}$ increases 
and becomes less dependent on temperature. In the extreme 
dilute region $Z^{F} \longrightarrow 1$. As far as $Z^{B}$ 
is concerned it follows the behaviour of $Z^{F}$ only in a close 
vicinity of $n_{tot}=2$. Going away from such case the nonbonding
spectral weight $Z^{B}$ decreases as a direct consequence
of the relations (\ref{ZB},\ref{pairs}).

\begin{figure}
\centerline{\epsfxsize=8cm \epsfbox{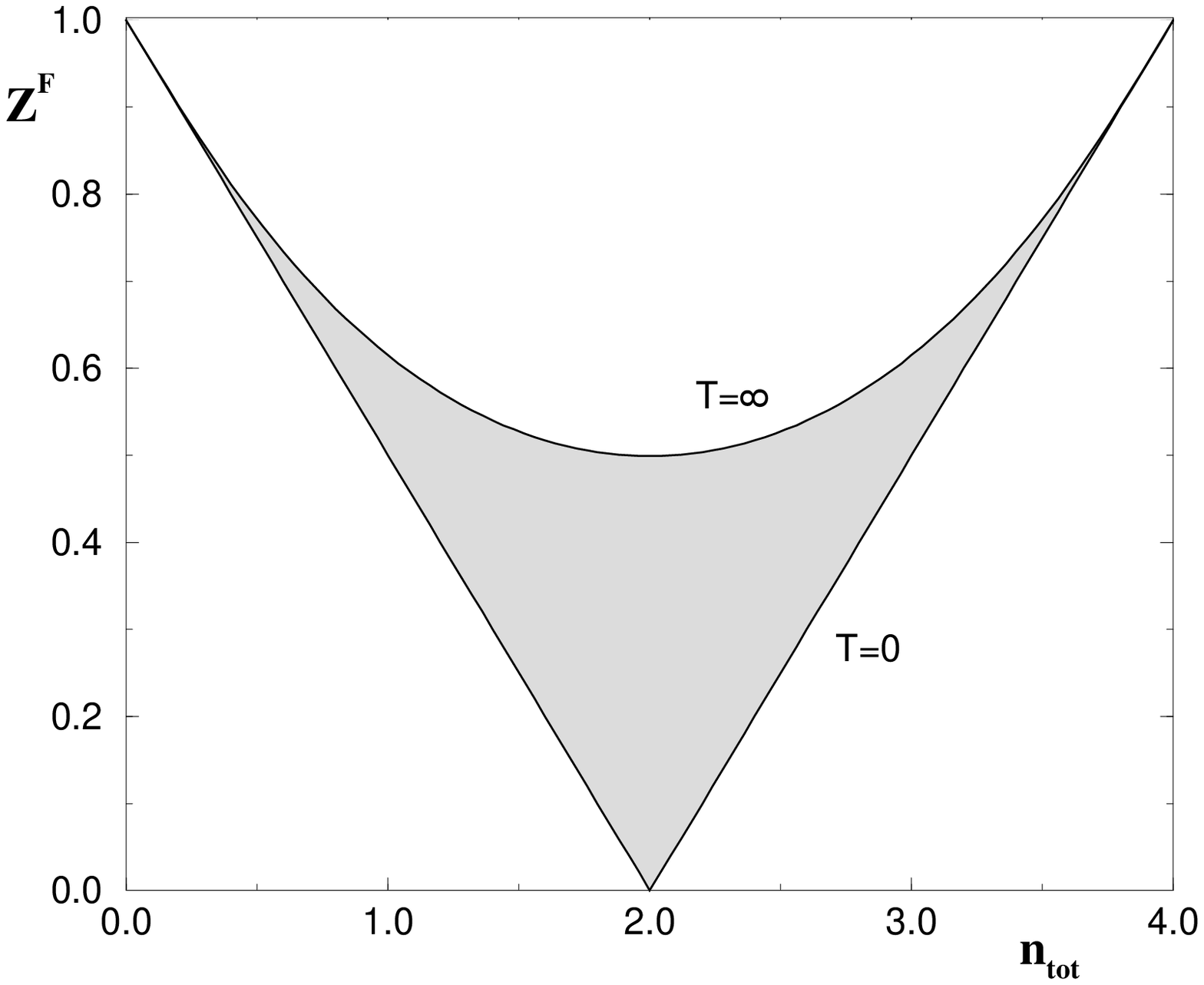}}
\centerline{\epsfxsize=8cm \epsfbox{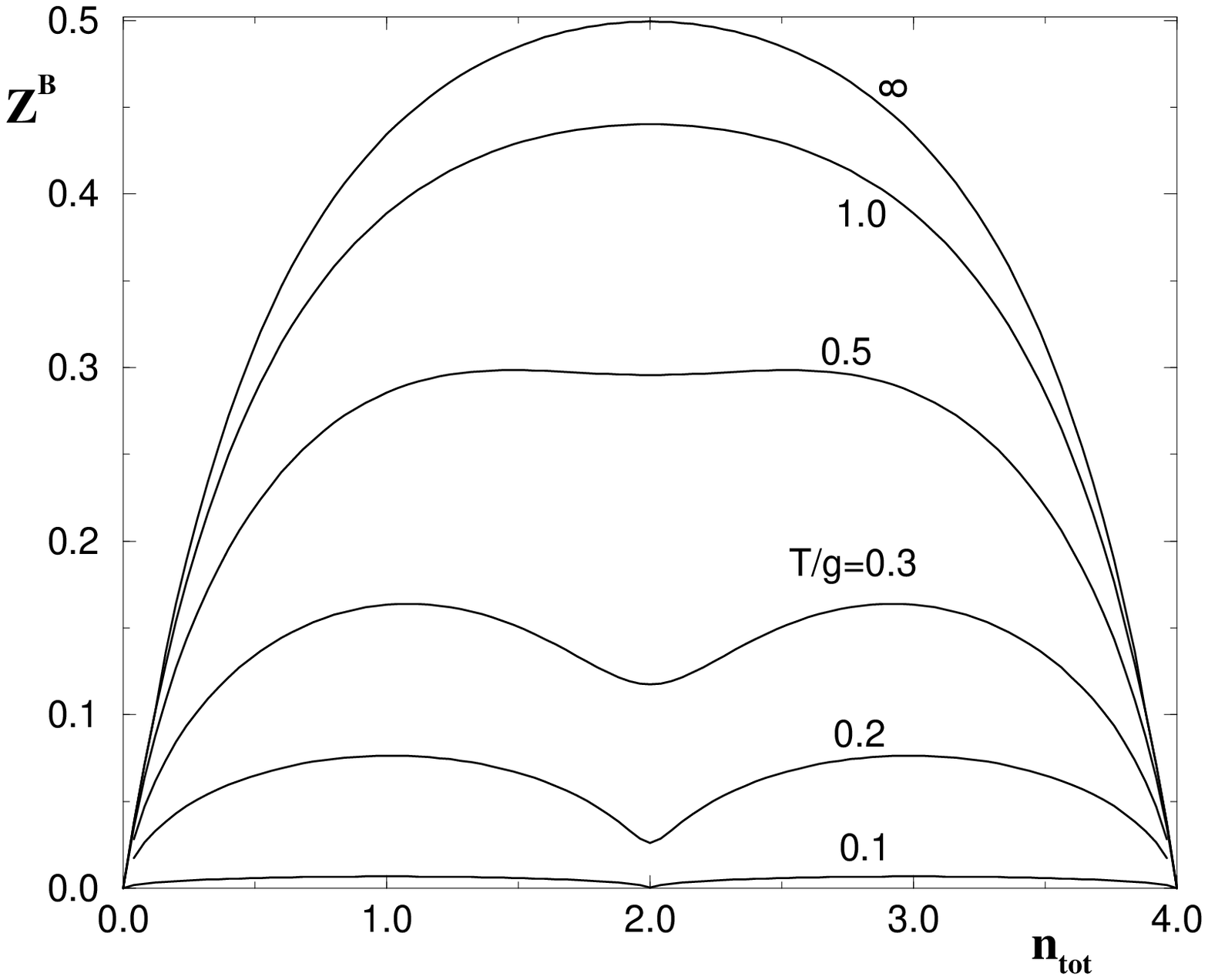}}
\caption{(a) Spectral weight of the nonbonding fermion state
as a function of total charge concentration $n_{tot}$ per site.
All finite temperature values are situated within the shaded
area in the figure. (b) Spectral weight of the nonbonding
state in the hard core boson subsystem for several
temperatures as indicated. Both figures (a) and (b)
are obtained for $\Delta_{B}/2=\varepsilon_{f}$.}
\end{figure}

Parameter $\Delta_{B}$ has rather a minor effect on both 
spectral weights, 
it mainly affects their temperature variation similarly to what 
is shown in figure 1. In order to characterise the temperature 
dependence of $Z^{F}$ we define characteristic temperature
$T^{*}(n_{tot},\Delta_{B})$ \cite{Domanski98} which is an 
inflexion point $d^{2} Z^{F}(T^{*})/dT^{2} = 0$. Roughly 
speaking, the spectral weight $Z^{F}$ starts to decrease 
when temperature drops below $T^{*}$.  From higher dimensional 
studies of the BF model in the symmetric case ($n_{tot}=2$, 
$\Delta_{B}=2\varepsilon_{f}$) \cite{Domanski98,romano} it 
is known that suppression of the nonbonding state below 
$T^{*}$ is accompanied by an appearance of the pseudogap 
structure. Apart of  the symmetric case there is not enough 
evidence that such relation remains valid.

Reduction of the nonbonding state spectral weight $Z^{F}$
for temperatures near and below $T^{*}$ is closely 
related to appearance of the pairing-type correlations. 
To prove this let us consider the Green's function $\langle 
\langle b_{i} c_{i,\downarrow}^{\dagger} ; c_{i,\uparrow}
^{\dagger} \rangle\rangle _{\omega}$ given in equation
(\ref{bcc_analytic}) which yields the following correlation 
function
\begin{equation}
\langle c_{i,\uparrow}^{\dagger} c_{i,\downarrow}^{\dagger}
b_i \rangle = g \left( 1 - Z^{F} \right) \; \frac{
f(\varepsilon_{+}) - f(\varepsilon_{-})}{\varepsilon_{+}
-\varepsilon_{-}}  \;.
\label{supercond_fluct}
\end{equation}
Let us recall that on a level of the mean field theory 
\cite{model,Micnas_02} the superconducting order parameter 
is given as
\begin{equation}
\langle c_{i,\uparrow}^{\dagger} c_{i,\downarrow}^{\dagger} 
\rangle = - g\; \langle b_{i} \rangle \sum_{\bf k} 
\frac{1}{2\tilde{\varepsilon}_{\bf k}} \; \mbox{tanh}
\left( \frac{\tilde{\varepsilon}_{\bf k}}{2k_{B}T} \right) ,
\end{equation}
where $\tilde{\varepsilon}_{\bf k}=\sqrt{(\varepsilon_{\bf k}
-\mu)^{2}+|g \langle b_{i} \rangle |^{2}}$ and $\varepsilon_{\bf k}$
denotes a dispersion of itinerant fermions. In the atomic limit
the order parameters are on average equal zero $\langle 
c_{i,\uparrow}^{\dagger} c_{i,\downarrow}^{\dagger}\rangle 
= 0 = \langle b_{i} \rangle$. We can think of a finite value 
(\ref{supercond_fluct}) as a result of fluctuating pairing
correlations. Magnitude of pairing correlations vanishes at
high temperatures while, for temperatures $T \leq T^{*}$,
achieves the finite value proportional to the spectral 
weight depleted from the nonbonding state ($1-Z^{F}$).
Figure 3 illustrates the temperature dependence of 
pairing correlations for several values of $\Delta_{B}$ 
at the fixed charge concentration $n_{tot}=2$.
Magnitude of $\langle c_{i,\uparrow}^{\dagger} c_{i,
\downarrow}^{\dagger} b_i \rangle $ turns out to be
proportional to the mean field value of $T_{c}^{MF}$ which 
proves their close relation.

\begin{figure}
\centerline{\epsfxsize=8.5cm \epsfbox{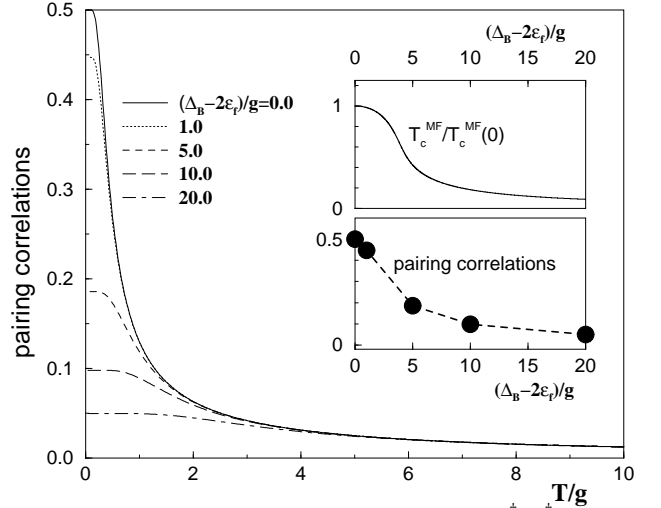}}
\caption{The pairing correlation function $\langle
c_{i,\uparrow}^{\dagger} c_{i,\downarrow}^{\dagger}
b_i \rangle $ induced in the atomic limit by the boson fermion
coupling $g$. The upper inset shows a mean field value
$T_{c}^{MF}$  for a fixed ratio $g/D=0.1$, where $D$ denotes
the fermion bandwidth. The lower inset is
the ground state value of the correlation function.
The main figure and the insets were obtained for
$n_{tot}=2$.}
\end{figure}     

The boson fermion model is claimed by some authors 
\cite{model,Tripodi_02,Micnas_02,Auerbach_02} to capture 
key aspects of the theory for high temperature 
superconductors (HTSC). In realistic description of 
the HTSC materials one must however consider their anisotropic 
$dim=2+\delta$ structure. Pairing correlations discussed 
here for the atomic limit would in higher dimensions lead
to: (a) formation of fermion pairs at $T_{p}$, and (b) 
at $T_{c} \leq T_{p}$ to their long range coherence, 
establishing the superconductivity (with $T_{c} \neq 0$). 
What remains to be studied for the realistic $dim=2+\delta$ 
systems is a pseudogap region of the incoherent fermion pairs 
$T_{c} \leq T \leq T_{p}$. We hope that the exact solution of 
the BF model discussed here for the atomic limit may help in 
such future investigations.

\end{multicols}

\begin{references}
\bibitem{model}
    J.\ Ranninger and S.\ Robaszkiewicz, Physica {\bf B}, 468 (1985); 
    R.\ Micnas, J.\ Ranninger and S.\ Ro\-basz\-kie\-wicz,
    Rev.\ Mod.\ Phys.\ {\bf 62}, 113 (1990); 
    J.\ Ranninger and J.M.\ Robin, Physica {\bf C 253}, 279 (1995). 
\bibitem{perturbative}
    J.\ Ranninger, J.M.\ Robin and M.\ Eschrig, 
    Phys.\ Rev.\ Lett.\ {\bf 74}, 4027 (1995); 
    J.\ Ranninger and J.M.\ Robin, Solid State Commun.\
    {\bf 98}, 559 (1996);
    J.\ Ranninger and J.M.\ Robin, Phys.\ Rev.\ B 
    {\bf 53}, R11961 (1996); 
    P.\ Devillard and J.\ Ranninger, Phys.\ Rev.\ Lett.\ 
    {\bf 84}, 5200 (2000).
\bibitem{Domanski98}
    T.\ Doma\'nski, J.\ Ranninger and J.M.\ Robin,
    Solid State Commun.\ {\bf 105}, 473 (1998).
\bibitem{romano}
    J.M.~Robin, A.~Romano, J.~Ranninger, 
    Phys.\ Rev.\ Lett.\ {\bf 81}, 2755 (1998); 
    A.~Romano and J.~Ranninger, 
    Phys.\ Rev.\ B {\bf 62}, 4066 (2000).
\bibitem{Domanski01}
   T.~Doma\'nski and J.~Ranninger, Phys.\ Rev.\ B 
   {\bf 63}, 134505 (2001).
\bibitem{Tripodi_02}
   L.~Tripodi and J.~Ranninger, Phys.\ Rev.\ B (2003)
   {\em in print}, cond-mat/0212332.
\bibitem{Micnas_02}
    R.\ Micnas, S.\ Robaszkiewicz and A.\ Bussmann-Holder, 
    Phys.\ Rev.\ B {\bf 66}, 104516 (2002); 
    Physica {\bf C 387}, 58 (2003).
\bibitem{Zubariev60}
    D.N.\ Zubariev, Sov.\ Phys.\ Uspekhi {\bf 3}, 320 (1960). 
\bibitem{Micnas92}
    R.\ Micnas and S.\ Robaszkiewicz, Phys.\ Rev.\ B
    {\bf 45}, 9900 (1992). 
\bibitem{Auerbach_02}
    E.\ Altman and A.\ Auerbach, Phys.\ Rev.\ B {\bf 65},
    104508 (2002); \\
     M.\ Mierzejewski and E.\ Kochetov, 
     cond-mat/0204420 (unpublished); 
     R.\ Friedberg, T.D.\ Lee and H.C.\ Ren, Phys.\ Rev.\ B
    {\bf 50}, 10190 (1994).
\end{references}
\end{document}